# Werner Heisenberg and the German Uranium Project (1939 – 1945)

# Myths and Facts

Klaus Gottstein, Munich, Germany, June 2016

## Summary

*A careful analysis of the available sources on Heisenberg's work and further activities during World War II and during his internment at Farm Hall after the end of the war leads to the following summary: "Heisenberg, like several other German physicists, was drafted by German Army Ordnance when war began in Europe in September 1939 to investigate whether the energy from splitting Uranium nuclei by neutrons could be used for military purposes. Heisenberg found that this is possible in principle but would require such enormous industrial expenditures that it would take many years and would be impracticable while the war lasted. The project was therefore dropped by the Nazi Government in 1942. Heisenberg had even refrained from calculating a precise value for the critical mass of U 235. He was relieved that he was thus spared a moral decision between obeying an order to build the bomb or risking his life by refusing to be involved in the project or sabotaging it. He was happy to be confined to a project of building a small test reactor under civilian administration that the Government had approved.. In 1941 Heisenberg had tried to get the opinion of Niels Bohr in Copenhagen on what the international community of nuclear physicists could possibly do or prevent regarding the long-range technical feasibility of making nuclear weapons. Bohr had misunderstood Heisenberg's cautious approach."*

## Contents

**Introductory Remarks**

In a recent review article of a French novel[1] in the literary supplement of a widely-read German daily newspaper[2] Heisenberg is called "a pioneer of the atom bomb". This is only another example of the many misunderstandings and misrepresentations which often appear in the media on various occasions and can even be found in some publications in scientific journals and books. The gist of the message in publications of this type is "Heisenberg tried to build an atom bomb for Hitler and failed".

The expression of such oversimplifying, misinformed and misleading opinions began already right after the end of war in Europe in 1945. The construction of atom bombs in the "Manhattan Project" had been started under the assumption that the Germans under the scientific leadership of Heisenberg were also beginning to be engaged in making atom bombs so that the physicists at Los Alamos felt, or at least suspected, that they were "in a race with Heisenberg". When it turned out on inspection of the laboratory of Heisenberg's team at Haigerloch by a U.S. advance team on 23 April 1945 that only a small test reactor had been constructed that had not reached criticality yet, the Allies were led to believe by their own prejudices that this was Heisenberg's inadequate approach to an atom bomb.

In order to avoid the cementation of prejudices and misconceptions in the historical memory of future generations it seems appropriate, as long as some of those are still alive who have known Heisenberg and many of his colleagues and collaborators personally[3], to summarize the evidence and straighten out some misperceptions that have arisen in some parts of the literature. That is the purpose of the present paper. The procedure chosen is to list a few obvious questions which arose during and after World War II. The attempt to find answers for these questions on the basis of incomplete, false or misinterpreted information led to the errors mentioned above. It will be tried to confront them with the facts, as far as these are known, and thereby contribute towards a better understanding of the difficult situation which one of the greatest physicists of the last century had to face under a dictatorial regime in time of war.

---

[1] „Le principe" by Jérome Ferrari. In this novel Heisenberg's thinking is described in a poetic way as being determined by a "principle of beauty". The reviewer finds this remarkable for a "pioneer of the atom bomb".
[2] Süddeutsche Zeitung, 26 February 2015, page 12
[3] Klaus Gottstein was a member of the Max Planck Institute for Physics from 1950 to 1970 under the directorship of Werner Heisenberg. For several years he was head of the experimental division of the Institute. In 1969 he asked Heisenberg about his visit to Bohr in 1941, and Heisenberg told him.

**Four Questions and Answers**

**Question 1:** How did Heisenberg get connected with investigations on the feasibility of using the energy freed by the splitting of uranium nuclei by bombardment with neutrons for driving machines or producing explosive weapons?

**Facts and comments:**

The war began on 1 September 1939, and a few weeks later Heisenberg, who was professor at the University of Leipzig, and some other German physicists were drafted by Army Ordnance to explore the feasibility of a nuclear bomb which, after the discovery of fission and of the chain reaction, could not be ruled out. How real was this theoretical possibility? Heisenberg was given the task to find out.

Army Ordnance also sequestered the Kaiser Wilhelm Institute for Physics in Berlin. Its director, Peter Debye, a Dutch citizen, was given the choice to accept German citizenship or resign for the duration of the war. He preferred the latter and emigrated to the United States. An official of Army Ordnance, the physicist Dr. Kurt Diebner, was appointed interim director of the institute.

**Question 2:** How far did Heisenberg pursue his investigations and what was their result?

**Facts and comments:**

By 1941 Heisenberg, after almost two years of intense theoretical and experimental investigations by the drafted group known as the "Uranium Club", had reached the conclusion that the construction of a nuclear bomb would be feasible in principle, by Uranium isotope separation or by Plutonium production in reactors, but both ways would take many years. They would be beyond the means of Germany in time of war, and probably also beyond the means of Germany's adversaries. This opinion was accepted by the leading authorities of the Nazi Government. (When Heisenberg heard about the Hiroshima bomb, almost four years later while interned at Farm Hall, at first he could not believe it was a nuclear bomb.) Munitions and Armaments Minister Speer offered support for a small reactor project as a possible source of electric power. Thus, the "military project" was abandoned as being useless for Germany's "final victory" in the present war. The sequestration of the Kaiser Wilhelm Institute for Physics was ended, Dr. Diebner returned to his facilities at Army Ordnance where he headed a separate group, and the institute returned to civilian administration by the Kaiser Wilhelm Society under the general direction of the civilian Reichsforschungsrat (Reich Research Council). Heisenberg was very happy with this outcome which spared him the moral decision whether to participate in a large bomb project or risk his life by refusing to cooperate in it. In July 1942 Heisenberg was appointed director **at** the Kaiser Wilhelm Institute for Physics (not **of** the Institute because this position remained reserved for Peter Debye).

Thus, Heisenberg's "failure" had nothing to do with "moral scruples", sabotage or incompetence, as has been suggested by some authors. The project had just been terminated because it had shown that "bomb-building" would be extremely expensive, lengthy and useless for winning the war. In a personal letter written to a friend in October of 1941 Heisenberg called it "imagination run wild" (Phantasterei) to think of the use of atomic energy for large-scale destruction though he did not exclude that for the distant future.[4] Nevertheless, Heisenberg's theoretical investigations, carried out by the end of 1939 and the beginning of 1940, allowed the possibility of technical use of the energy released by splitting uranium nuclei. Experiments done at Leipzig and in other German laboratories showed that a "Uranium Machine" with natural Uranium and heavy water could function.

In the course of his work Heisenberg commuted between Leipzig and the Kaiser Wilhelm Institute for Physics in Berlin. In 1942 Heisenberg moved to Berlin for preparations for the experimental Uranium reactor.

**Question 3:** What was the purpose of Heisenberg's call on Bohr in Copenhagen in 1941?

**Facts and comments:**

Although by the second half of 1941 Heisenberg was convinced that in the next few years the construction of a nuclear bomb was not feasible, the question remained: What about the long-range future? Elisabeth Heisenberg reports in her book[5] that her husband tortured himself with the thought that in the long run the

---

[4] Letter of October 1, 1941 by Werner Heisenberg to Hermann Heimpel, quoted in the brief outline of the history of the German Uranium project and of Heisenberg's activities during the war in the Introduction by Helmut Rechenberg to Heisenberg's „Ordnung der Wirklichkeit", page 17, i.e. to the "Manuscript of 1942", see Literature at the end of this article

[5] Elisabeth Heisenberg, Das politische Leben eines Unpolitischen. Erinnerungen an Werner Heisenberg, R. Piper & Co Verlag München 1980

United States with her superior industrial capacity might be able to produce atom bombs and use them on Germany if war conditions dragged on long enough. But was the final construction of nuclear weapons unavoidable? Was it conceivable that the then small international community of nuclear physicists could arrive at an agreement to refrain from the construction of these entirely new weapons of mass destruction?

Ever since working with Bohr in Copenhagen in the 1920s Heisenberg had been used to discussing with his friend and mentor Bohr difficult questions which arose in the course of their work. It was suggesting itself that also in this case it would be helpful to discuss the matter with Bohr and get his opinion. What Heisenberg, in a kind of naiveté, did not realize was that his old cordial relationship with Bohr had been affected by the events of the war. For Bohr his old friend Heisenberg was now a representative of an enemy country, of the occupying power of his native Denmark, whose remarks would have to be looked upon with suspicion. Heisenberg managed to make the trip to Copenhagen in September of 1941, using the opportunity of a conference on astrophysics arranged by the German Culture Institute in Copenhagen. Bohr boycotted this Institute set up by the German Foreign Ministry for propaganda purposes after the occupation of Denmark. For Heisenberg accepting an invitation to lecture at the Institute was a means to obtain a visa for a visit to Copenhagen that would have been unobtainable otherwise. It also provided an opportunity to call on his old friend Bohr in an unobtrusive way.

Heisenberg spent several days in Copenhagen and probably saw Bohr several times, in Bohr's office, in Bohr's home and on a walk. On the latter occasion when there was no danger of being overheard by the Gestapo, Heisenberg undertook to broach the questions which were the real reasons for his trip. He was extremely cautious in choosing his language. Mentioning to Bohr the existence of a German nuclear programme and of his participation in it, could be interpreted as, and probably was, treason punishable by death. So he used very involved expressions which, he assumed, Bohr would understand but which to uninitiated Gestapo agents, if they heard of them later by some incautiousness, could be explained away as harmless conversation. This is what Heisenberg told the present author (K. G.). He regretted after the war that he had not been more straightforward, in spite of the risks involved. His intended mission foundered. As soon as Bohr understood that Heisenberg was beginning to talk, though indirectly, about his assured knowledge that nuclear bombs were feasible in principle, Bohr broke off the conversation and would not hear any more about this subject. He could not imagine that Heisenberg acted on his own initiative, without any special permission, let alone orders, by German authorities. But this was so. Heisenberg had thought, naively, as mentioned above, that Bohr would be ready, as he always had been in earlier times, to discuss with him possible solutions for complicated problems. He had lacked the sensitivity for Bohr's patriotic feelings under the changed circumstances of war and occupation. On the other hand, it is justified to say that it took great moral courage for Heisenberg to talk to Bohr about implications of his secret work. Heisenberg risked his neck.

Bohr, however, had looked with misgivings at the motives of Heisenberg's visit under the conditions of German occupation of Denmark. Bohr was, at the time of the visit in 1941, distressed by the circumstances of Heisenberg's visit, his lecture at the German Culture Institute and his contacts with the German Embassy (more correct: Legation) in Copenhagen.[6] For Bohr it was of central and sad significance that Heisenberg during his visit expressed his conviction of a final German victory whereas Bohr, as a Danish patriot, had placed all his hopes in a German defeat. In September of 1941, with large parts of Europe occupied by Germany, German troops approaching Moscow, and the United States continuing to remain neutral, Heisenberg concluded that Germany might win the war after all. At the beginning of the war he had, in private, expressed the view that Hitler would lose the war like a chess-player would lose a game into which he entered with one castle less than his opponent. But now Heisenberg like most non-Nazi Germans had come to the conclusion that a German victory seemed likely. They feared that a German defeat would mean Soviet occupation of Europe which, even for anti-Nazis, was considered an even greater evil than German domination. Auschwitz and the full extent of Nazi crimes were not yet known, but Stalin's massacres were. The hope - completely unrealistic as we now know but considered realistic at the time - was that after a German victory the German army would get rid of Hitler and his henchmen. The anti-Nazi stance of many German generals, who later took part in the assassination plot of July 20, 1944, was known to persons who, like Heisenberg through the "Wednesday Society", were close to opposition circles. For Heisenberg, it was part of his care for Bohr to think in sober terms of the future and of Bohr's and his institute's survival. It would be advisable to end opposition to a victorious Germany, Heisenberg suggested to Bohr. It would be better for Bohr and his institute, Heisenberg felt, to have normal relations with the German Legation in Copenhagen. He knew that some of its diplomats were non-Nazi and ready to assist Bohr in any way at their disposal. (One of these diplomats, Georg Ferdinand Duckwitz, later informed the Danish underground movement of the impending arrest and deportation of Danish Jews. This led to

---

[6] This account of how Heisenberg opened his conversation with Bohr and how Bohr reacted is based on a report by Bohr to Eugen Feinberg when Bohr visited Moscow in May 1961. Further details are given on the ensuing pages.

the rescue of the Danish Jews by their escape to Sweden. After the war Duckwitz was Ambassador of the Federal Republic of Germany in Copenhagen.) But for Bohr who as a Danish patriot steadfastly refused to have anything to do with German authorities, Heisenberg's well-meant suggestion sounded like an invitation to collaboration with the Germans. He even suspected that Heisenberg, had their conversation continued, would have tried to persuade him to take part in his work on a German atomic bomb

It is often claimed in the literature that Heisenberg's aborted conversation with Bohr in Copenhagen in 1941 was the end of their personal friendship. This is not true. Still in Copenhagen, before his return to Germany, Heisenberg wrote a letter to his wife Elisabeth the recent discovery of which caused much excitement. It has been published[7] in Heisenberg's collected letters (1937 – 1946) to his wife. This letter shows that, two days after his famous, misunderstood conversation with Bohr, Heisenberg spent a harmonious evening with Bohr at Bohr's home. They discussed physics, Heisenberg played the piano, and Bohr read a story to him. Thus, Bohr's "rage" after the ill-fated discussion cannot have been as deep as is often assumed. Their personal friendship continued, as is also shown by the fact that they visited each other after the war with their families in their homes and spent their vacations together in Greece or South Italy, and that Bohr wrote an article for the Festschrift to Heisenberg's sixtieth birthday in 1961.

**Robert Jungk's Book**

Another serious trial of their friendship arose, however, as the American-Austrian, German-born journalist Robert Jungk published, in 1956, his bestseller "Heller als tausend Sonnen"[8] on the construction of the atomic bomb in the United States and on the nuclear work in Germany during the war. He had interviewed many of the leading physicists in both countries. Heisenberg and Carl Friedrich von Weizsäcker had freely cooperated with him. They had told Jungk about the intended purpose of the failed mission to Copenhagen in September of 1941. Weizsäcker, with several other German scientists, had also attended the astrophysics conference in the German Culture Institute but had not been present at Heisenberg's conversation with Bohr. But Heisenberg had informed Weizsäcker about that failure immediately afterwards, and Weizsäcker had supplemented Heisenberg's report to Jungk.

---

[7] Werner Heisenberg, Elisabeth Heisenberg, Meine liebe Li, Der Briefwechsel 1937 – 1946, Herausgeber Anna Maria Hirsch-Heisenberg, Residenz Verlag Salzburg 2011
[8] Robert Jungk, Heller als tausend Sonnen. Das Schicksal der Atomforscher, Alfred Scherz Verlag Bern 1956

In his book, however, Jungk embellished the sober descriptions he had received from the two German physicists by interpretations created by his own imagination. Thus, he presented Heisenberg's satisfaction with the technical difficulties of bomb construction and his lack of enthusiasm for overcoming these difficulties, as a secret plan to prevent, for moral reasons, the construction of an atomic bomb for Hitler which otherwise he could have built. Heisenberg, and particularly von Weizsäcker, wrote long letters to Robert Jungk in which, while appreciating Jungk's extensive research and detailed accounts of the developments, criticized some of his generalisations and exaggerations. Cathryn Carson, in her article "Reflexionen zu 'Kopenhagen'", appended to the German edition of Frayn's play "Copenhagen", quotes from these letters[9]. In the Danish translation of his book, which appeared in 1957, Jungk published an extraction of Heisenberg's letter, but only the laudatory part. He omitted the criticisms and also Heisenberg's remark in his letter that he would not like to be misunderstood as having exerted resistance against Hitler. These omittances are particularly regrettable because in some quarters it was even assumed that Heisenberg had commissioned Jungk's book. This did much to harm Heisenberg's credibility. Heisenberg never "portrayed himself after World War II as a kind of scientific resistance hero who sabotaged Hitler's efforts to build a nuclear weapon", as was suggested, e.g., by James Glanz in The New York Times of February 7, 2002 after the publication of Bohr's unsent letters to Heisenberg (see below). On the contrary, Heisenberg always stressed how content he had been that nuclear weapons did not seem to be feasible for several years to come so that Hitler and his government, when this had become clear to them, made no effort to build them. Bohr read the book in the Danish edition and took exception to Jungk's description of his 1941 meeting with Heisenberg. This is understandable because Jungk described as completed conversation what Heisenberg had intended to discuss with Bohr but had not got a chance to ventilate because of Bohr's refusal to listen to Heisenberg's involved nuclear hints. Bohr, however, was led to believe that Heisenberg had authorized Jungk's description. But Bohr did not object in public to Jungk's presentation. He just drafted a letter to Heisenberg which he never posted.

**Bohr's Unsent Letter to Heisenberg**

When it became known that the Niels Bohr Archive in Copenhagen held a letter by Bohr to Heisenberg, written after the appearance of Jungk's book but never

---

[9] Carson, Cathryn. "Reflections on Copenhagen . In: Michael Frayn's Copenhagen in debate: Historical essays and documents on the 1941 meeting between Niels Bohr and Werner Heisenberg, ed. Matthias Dörries, Berkeley: Office for History of Science and Technology, 2005. Published in German as "Reflexionen zu 'Kopenhagen.'" In: Michael Frayn, Kopenhagen: Mit zehn wissenschaftsgeschichtlichen Kommentaren, ed. Matthias Dörries, 3rd, rev.ed., p. 172-188. Göttingen: Wallstein, 2003. In initial form in 1st ed., p. 149-162. Göttingen: Wallstein, 2001

sent, speculation concentrated on this document from which some observers expected the solution of all the open questions. It was to be published 50 years after Bohr's death, i.e. in 2012. However, to end speculation, the Niels Bohr Archive, around early 2002, released 11 documents pertaining to Heisenberg's visit, including the much-discussed unsent letter, preceded by an article by Aage Bohr, first published in 1967, on "The War Years and the Prospects Raised by Atomic Weapons". The documents, with the exception of one letter written by Heisenberg to Bohr, are unfinished drafts written by Bohr in the late 1950s and early 1960s, addressed to Heisenberg, but never sent. As the director of the Niels Bohr Archive, Finn Aaserud, points out, the documents have to be viewed with caution. They were written 16 years or more after the event and represent just drafts, not finished papers. Nevertheless, the contents of the documents are interesting and, depending on the pre-established views and opinions of the readers of today, surprising to a lesser or greater degree. Here are some of the general characteristics of the documents:

- Bohr's tone in addressing Heisenberg is extremely cordial and friendly.
- Bohr was still highly interested in clarifying Heisenberg's intentions and motivations behind his 1941 visit. His sentences in Document 11 c "I have long been meaning to write to you ..." and "I have written in such length to make the case as clear as I can for you and hope we can talk in greater detail about this when opportunity arises" are proof of this. (This is new information. Heisenberg was under the impression that Bohr and he, having differing recollections of their discussion, had come to the conclusion that it would be best to let rest the spirits of the past. It is a pity that the letter was not sent. Several opportunities for clarifying conversations were missed at later meetings of Bohr and Heisenberg. It seems that Bohr was afraid he might hurt Heisenberg's feelings by insisting too much on his interpretation of the events.)
- Document 1 contains the confirmation that Bohr and Heisenberg met several times during Heisenberg's visit to Copenhagen in 1941: Bohr refers to "our conversations" in the plural, and he mentions "our conversation in my room at the institute" as well as the strong impression Heisenberg's remarks made "on Margrethe and me". Since it is unlikely that Bohr's wife Margrethe was present at the confidential conversation in Bohr's room in the institute one may assume that Heisenberg's recollection is correct that he was also invited to Bohr's home. This is confirmed by Heisenberg's much later discovered letter to his wife written before his departure from Copenhagen in 1941. Moreover, there is Heisenberg's and von Weizsäcker's testimony that the critical discussion took place during a walk, to avoid unwanted earwitnesses.
- Bohr understood and appreciated that one of Heisenberg's reasons for the visit was genuine care: to see how Bohr and his institute fared under German occupation and to be of assistance, if at all possible (Document 11 c). Bohr suspected, however, that the main reason for Heisenberg's visit was to get him, Bohr, involved in Germany's atomic bomb project which, Bohr thought Heisenberg had cautiously hinted, existed in reality under his, Heisenberg's, leadership. When Bohr came to this conclusion he stopped the conversation. This is admitted by Bohr in Document 11 c where he writes "During the conversation, which because of my cautious attitude was only brief ...". Bohr's cautiousness was justified by his fear that any words he might speak would be somehow made known to German authorities. There is no indication of an awareness by Bohr that Heisenberg was under the same handicap. In public conversations, also in the cafeteria of Bohr's institute, he may have had to say things which did not represent his true opinion. (This situation is well-known to people having lived under cruel dictatorships.)

- Document 6 says that Heisenberg "did not wish to enter into technical details but that Bohr should understand that he knew what he was talking about as he had spent 2 years working exclusively on this question." Bohr had known about the possibility of nuclear weapons only in a very general way and at that time still had held the opinion that the technical difficulties were insurmountable. Bohr had been "doubtful looking" (Document11a). Therefore Heisenberg found it necessary to mention his two years of investigations in order to convince Bohr that he was not "talking moonshine". Bohr interpreted this, erroneously, as meaning that Germany was working, with Heisenberg's leading participation, on the production of atomic bombs. As mentioned above, Bohr could not imagine that Heisenberg would reveal a state secret of this importance to him, a foreigner, unless he was authorized, or even ordered, to do so. But Heisenberg's trust in Bohr was of such depth that he dared to do that on his own initiative though very cautiously. To his dismay, Bohr did not allow him to complete his cautious message that the construction of an atomic bomb would take several years so that it would NOT be attempted in Germany for the near future. The question for which Heisenberg would have liked to know Bohr's opinion was whether it might be possible to come to an agreement within the still relatively small international community of nuclear physicists not to work on the construction of atomic bombs at all. This, for Heisenberg, was the central reason for his trip to Copenhagen in 1941 and his visit with Bohr.

For Bohr, however, the all-important message was Heisenberg's advice to take into account the imminent victory of Germany, anyhow the lesser evil compared to Soviet occupation, and stop boycotting German institutions in Copenhagen. For Heisenberg this had been only an introductory item of secondary urgency for opening the conversation. Bohr saw under this advice also Heisenberg's hidden reference to atomic bombs and interpreted it as an indirect, most unwelcome invitation for cooperation also in this area. For Heisenberg his advice to prepare for an apparently unavoidable victory of Germany was of secondary importance compared to the question how to deal with the "open road" to atomic armaments. Therefore this aspect of his 1941 visit to Copenhagen did not receive much attention in his reports after the war when this episode came up. For Heisenberg they were only of marginal importance so that he did not even mention them in the interview with Robert Jungk.

- The further development of the war in 1942 and later must have removed Heisenberg's conviction of 1941 that Germany will be victorious. Bohr wondered for many years whether this was the reason why Heisenberg, in retrospect, had forgotten or repressed these statements which Bohr clearly remembered but Heisenberg, according to Robert Jungk, did not mention. On the other hand, again according to Robert Jungk, Heisenberg claimed to have said things to Bohr which Bohr was sure not to have heard. These concerned Heisenberg's questions about Bohr's opinion regarding an international agreement of the community of nuclear physicists not to make atomic bombs. Bohr did not know that Heisenberg, in a letter to Jungk, clarified that he had indeed planned to ask Bohr questions of this kind but was given no chance to ask them because Bohr ended the conversation abruptly when the topic of atomic bombs was touched. Jungk had presented as established fact what Heisenberg had just intended to do. Thus, it was not surprising that Bohr did not remember what Jungk described.

**Feinberg's Report on Bohr's Reminiscences**

E. L. Feinberg reports in his book *Physicists. Epoch and Personalities*[10] that Bohr, when he visited Moscow in May 1961, 16 months before his death, was still pondering about the possible reason for the discrepancy between what he remembered about the conversation with Heisenberg in Copenhagen in 1941 and what Heisenberg, according to Jungk, seemed to remember. Bohr told Feinberg and other Russian listeners: „Heisenberg is a very honest man. It is astonishing, however, how one is capable of forgetting one's views if he is gradually changing them"[11] This appraisal of Heisenberg's character by Bohr agrees with a statement by Edward Teller in his memoirs, also reported by Feinberg[12]: "Heisenberg was not only a brilliant physicist but also a person whose decency and feeling of responsibility I could observe many times. I cannot imagine that he supported Nazis by his own good will, even less that he did it with enthusiasm as Bohr's version

---

[10] E. L. Feinberg, Physicists. Epoch and Personalities, World Scientific Publishing C., New Jersey London 2011, Sections 8.1 and 8.2 ("Tragedy of Heisenberg" and "Bohr and Heisenberg")

[11] E. L. Feinberg, loc. cit., page 298
[12] E. L. Feinberg, loc. cit., page 310

declares. How could it happen that Bohr misunderstood him? Information that I have gathered leads me to the thought that Heisenberg went to Bohr for moral advice … "

In other words: Bohr could not imagine that Heisenberg, an honest man, after the war deliberately distorted the truth when he reported his version of what was said during the famous 1941 meeting. If that differed from what Bohr thought he clearly remembered, the solution must be sought in psychology: Heisenberg's changing views on the outcome of World War II since 1941 must have changed, unconsciously, his memory of what he had told Bohr.

In reality, both were right. As Hans Bethe put it[13]: "The two famous physicists just talked past each other, starting from different assumptions." Each of them just remembered those parts of the conversation which concerned what he had considered to be the most relevant topic: Bohr his assumed invitation by Heisenberg to be involved in his suspected atomic bomb project, Heisenberg his failed attempt to get Bohrs opinion on what the international community of nuclear physicists might do regarding the road leading to atomic weapons, still closed at present for practical reasons but clearly open in years to come.

In this context ít may be of interest what Edward Teller told the present author (K. G.) during the lunch break at a conference in the U.S. in 1980[14], when what Teller called „the tragedy" of the misunderstanding between Bohr and Heisenberg during Heisenberg's visit to Copenhagen in 1941 came up in the conversation. Teller said that Bohr had been shocked and dismayed about the Nazis, was himself in a personally endangered situation, and apparently did not listen carefully enough. For any other person this would have been excusable, Teller said, but not for Bohr who had spent his life teaching complementarity and the necessity to use imperfect language for expressing the truth. Therefore he, Teller, would assign to Bohr the main responsibility for that tragical misunderstanding.

But also Heisenberg may have made a mistake by being too cautious. His friend and collaborator Carl Friedrich von Weizsäcker who had accompanied Heisenberg to Copenhagen and had waited in their hotel for the result of Heisenberg's conversation with Bohr and learned first-hand from the desperate Heisenberg about the complete failure of his mission thought later, as quoted by Feinberg[15] that Heisenberg approached his main topic much too slowly. He should have said immediately: "Dear Niels Bohr, I shall now tell you something which will cost my life if the wrong people learn about it. We study atomic weapons. It would be vital for humanity if we and our colleagues in the West would understand: All of us must work in such a way that a bomb will not be produced. – Do you think that might be possible?" Heisenberg talked too long in involved language and thus gave Bohr the chance to misunderstand and end the conversation before Heisenberg had completed his message. Jungk, however, in his book gave the impression that Heisenberg had been able to tell Bohr about the mere possibility of atomic weapons and that this shocked Bohr so deeply that he had become unable to listen any further.

---

[13] Hans A. Bethe, PHYSICS TODAY, issue of July 2000
[14] International Conference „A Global View of Energy", Miami, Florida, 1980

Heisenberg had not anticipated that Bohr would wonder who had authorized or ordered him to discuss with him military secrets. He had no strategy for dispelling suspicions of this kind. He had just assumed Bohr would understand that he spoke in his private capacity as Bohr's old friend and colleague who, however, because of the delicacy of the subject discussed, had to use very involved language. This assumption was sadly disappointed. Heisenberg was always sad that Bohr had misunderstood the purpose of his 1941 visit, and the unsent Bohr letters by Bohr also show that Bohr, unknown to Heisenberg, continued to ponder about the "mystery" why he and Heisenberg had so different memories of that event. (The "mystery", of course, was to a large extent Robert Jungk's doing.) In any case, probably because both of them thought that the other one preferred not to discuss the matter any further, they never tried to clarify their mutual misunderstanding.

Later during the war repeated German propaganda talks of the imminent use of "new weapons" fortified suspicions by Bohr and his Danish colleagues that there was a German nuclear bomb programme. Assertions to the contrary by Jensen, who visited Bohr a year later, were not trusted though he himself was considered honest. But Jensen was working on the reactor programme, and it had to be doubted that he was privy to all aspects of the programme.

- After Bohr's escape to Sweden and subsequent flight to Great Britain in the autumn of 1943 "it was quite clear already then, on the basis of intelligence reports, that there was no possibility of carrying out such a large undertaking in Germany before the end of the war". (Document 11 b). This is a remarkable confirmation of Heisenberg's own conclusion. It is also interesting that these intelligence reports had no influence on the progress of the Manhattan project.

---

[15] Eugen Feinberg, Werner Heisenberg – Die Tragödie des Wissenschaftlers. In: Werner Heisenberg by Hans-Peter Dürr, Eugen Feinberg, Bartel Leendert van der Waerden, Carl-Friedrich von Weizsäcker, Carl Hanser Verlag, München Wien 1977, 1992, pages 62, 63

**Question 4:** What did Heisenberg and von Weizsäcker mean to say at Farm Hall after hearing of the Hiroshima bomb?

Six months, from July 3, 1945 to January 3, 1946, ten German nuclear physicists and nuclear chemists (Erich Bagge, Kurt Diebner, Walther Gerlach, Otto Hahn, Paul Harteck, Werner Heisenberg, Horst Korsching, Max von Laue, Carl Friedrich von Weizsäcker, Karl Wirtz) were interned under comfortable conditions at "Farm Hall", a country mansion not far from Cambridge, used by the British Secret Service for the instruction of agents. Secret microphones were installed in their rooms, and their conversations were monitored, registered on coated discs and, as far as considered relevant, translated into English for the use of General Leslie Groves, the head of the U. S. Manhattan Project. In his memoirs "Now it can be told. The Story of the Manhattan Project", published in 1962, Groves revealed that these transcriptions existed in the archives. They were kept secret until 1991 when they were finally made available to historians and the interested public. The German originals had been deleted because the coated discs had been re-used after transcription.

The evaluation of the published English transcriptions resulted in a lively and often controversial debate in public discussions and writings on how to interpret the comments made by the internees, and in particular by Werner Heisenberg, Carl Friedrich von Weizsäcker and Otto Hahn, when they received the news of the atomic bomb on Hiroshima on August 6, 1945, and thereafter. Special attention was given to the following symptomatic remarks.

### Heisenberg's Contradictory Statements on the Order of Magnitude of the Critical Mass

1. Heisenberg's first reaction to the news heard at 6 p. m. on August 6 was that he did not believe that the atomic bomb mentioned was a true nuclear bomb. As justification for his disbelief he said that he could not imagine that the Americans had been able to procure the necessary two tons of uranium 235. Hans A. Bethe[16] concludes from this remark that Heisenberg cannot have worked on making nuclear weapons because in 1945 he still upheld for the critical mass of uranium 235 the old, much too large value of several tons that had been discussed before the war on the basis of simple random walk theories. Obviously he had not been interested in obtaining a precise value for the all-important critical mass. Bethe thought, as reported by Feinberg[17], that perhaps Heisenberg did not want to know. When asked about the critical mass, depending on the occasion, he had mentioned different values from ten kilograms to several tons. One example is the reply by Otto Hahn to Heisenberg's spontaneous reaction on August 6: "*But tell me why you used to tell me that one needed 50 kilograms of '235' in order to do anything. Now you say one needs two tons.*" Heisenberg gave an evasive reply. Another example is Heisenberg's reply at the Harnack House meeting on June 4, 1942 to Field Marshal Erhard Milch who had asked how large a bomb would have to be that could destroy a large city like London. Heisenberg is reported to have answered "about the size of a pineapple" which is not far from the truth if only the content of U 235 is meant. An anonymous report in spring 1942 to German Army Ordnance estimates for the critical mass a value between 10 and 100 kilograms. It is assumed that this estimate is due to Heisenberg. Manfred von Ardenne, German physicist and inventor and head of a private electronic and nuclear laboratory in Berlin recalls in his memoirs that Heisenberg told him around 1942 that only a few kilograms of U 235 would suffice for starting a chain reaction.

---

[16] H. A. Bethe, The German Uranium Project, *Physics Today, July 2000*

Only three hours later, at 9 p. m. on August 5, 1945, another radio announcement made it clear that the bomb dropped on Hiroshima was a Uranium bomb. Nine days later, on August 14, Heisenberg gave a lecture to his fellow internees in which he presented a correct theory of the atomic bomb. It showed that he would have been able to develop the correct theory of nuclear weapons, had he concentrated on that subject earlier.

**Carl H. Meyer's Analysis and Hypothesis**

Of particular interest in this context are the investigations by Carl H. Meyer and Günter Schwarz.[18] Puzzled by the contradictory reports on Heisenberg's ignorance or knowledge of the critical mass of U 235 they followed in detail the calculations made by Heisenberg in 1939/1940 on orders by Army Ordnance. After the war these calculations were published in Heisenberg's collected works (edited by W. Blum, H. P. Dürr and H. Rechenberg). Meyer and Schwarz thoroughly analyzed them. They find, as explained in their preprint (see footnote 18), that Heisenberg closed his calculations without giving a number for the critical mass. For whatever reason he did not take this last step which was within reach for him. This corroborates what Bethe concluded: Heisenberg did not want to know the correct value of the critical mass. He was interested in building a nuclear reactor, not nuclear weapons.

---

[17] E. L. Feinberg, loc. cit. Page 322
[18] Carl H. Meyer and S. Günter Schwarz, The Theory of Nuclear Explosives That Heisenberg Did not Present to the German Military, Preprint 467, Max Planck Institute for the History of Science, Berlin 2015, www.mpiwg-berlin.mpg.de/en/resources/preprints.html.

There remains the question why Heisenberg, for explaining his doubt about the nuclear character of the Hiroshima bomb, when hearing about it at 6 p. m. mentioned a ton value for the critical mass rather than one of the much smaller estimates which he had given in 1942 to Otto Hahn, Field Marshal Milch, Manfred von Ardenne and Army Ordnance which, though smaller, had still been considered to be much too high for technical realization in less than several years.

To this question Meyer and Schwarz offer quite a new hypothetical answer. Heisenberg made his outdated two-ton remark at the time between 6 p. m. and 9 p. m. when it was still uncertain and, for Heisenberg, even unlikely that the U. S. had accomplished the construction of a nuclear bomb. Heisenberg still held the belief that he and his team were further advanced in their work and their knowledge in this field than his American and British competitors and would be able to use this advanced knowledge as a "bargaining chip" in future negotiations. In this belief Heisenberg had been strengthened by his former close friend Samuel Goudsmit, in whose house Heisenberg had stayed when last visiting the United States in the summer of 1939. Goudsmit was now the scientific head of the ALSOS mission and had interrogated Heisenberg in Heidelberg in May 1945 after his arrest in Urfeld. When Heisenberg, naively but trustfully, had asked him about nuclear work in the U. S. since 1939 Goudsmit had replied that not much had happened because during the war U. S. physicists had other things to do. In the presence of British Major Rittner and possible further earwitnesses Heisenberg did not want to give his bargaining chip away by showing, prematurely, his advanced knowledge gained in two years of relevant studies. To satisfy the chemist Otto Hahn with a plausible reason why it was unlikely that the Americans had amassed enough U 235 for making a nuclear bomb the old, outdated ton-value for the critical mass would do. Unexpectedly, however, Hahn did remember the value of 50 kilograms that Heisenberg had estimated three or four years ago. This is the scenario Carl H. Meyer considered realistic for explaining the otherwise perhaps surprising figure Heisenberg had spontaneously available when suddenly confronted with the radio news of a so-called atomic bomb dropped on a Japanese city.

Another, much less sophisticated explanation for Heisenberg's return to an obsolete estimate for the critical mass would be that Heisenberg, for the last three years, had just worked on reactor construction, cosmic ray physics, elementary particle physics, philosophy and was happy not to have to think any longer about nuclear weapons and the critical mass. Anyway, its value he had never tried to calculate precisely. He had just roughly estimated it and by now (1945) it had escaped his memory or, at least, was not immediately available. The present author (K. G.) thinks that this simpler solution is probably the correct one though Meyer's scenario cannot be ruled out. Meyer deems it impossible that Heisenberg ever could have forgotten the order of magnitude of a natural constant of such fundamental importance as the critical mass of U 235!

Unfortunately, the U. S. mathematician and cryptologist Carl H. Meyer died just after the publication of the preprint mentioned in footnote 18 and in the list of Literature attached at the end. He had begun to write an extensive book on the life and work of Werner Heisenberg which now remains unfinished.

**Farm Hall Remarks by Carl Friedrich von Weizsäcker**

2.  According to the transcriptions, Carl Friedrich von Weizsäcker (CFvW) also made some spontaneous remarks upon hearing of the atom bomb dropped on a Japanese city by the U.S. The views or intentions that led to these remarks are often ascribed in the literature to Heisenberg's team as a whole, in particular because CFvW is known to have been a close collaborator and friend of Heisenberg. In these remarks CFvW expressed the following views:

    a. *Our team did not really want to make the bomb. Had we worked with the same devotion and intensity as the Americans we, too, could have succeeded.*
    To the second sentence Otto Hahn replied: "I do not believe that. And I am very happy that we did not succeed."
    The first sentence, however, seems to be supported also by an observation made in Russia during the last years of the war, as reported by E. L. Feinberg to CFvW in Moscow in 1987[19]: A thorough check of U.S. journals had shown that all the physicists in the U.S. who were considered capable of working on nuclear bombs had ceased

---

- [19] Carl Friedrich von Weizsäcker, Bewußtseinswandel, Carl Hanser Verlag, München Wien 1988, Kapitel 5 („Die Atomwaffe", Interview mit H. Jaenecke vom *Stern*, 1984) und Kapitel 6 („Nachtrag zum Gespräch zwischen Niels Bohr und Werner Heisenberg 1941"), pages 382/383 publishing. Apparently they were fully absorbed by secret work. Heisenberg, on the other hand, had published, as editor, a small volume on cosmic radiation which contained lectures by himself, CFvW, Wirtz and other members of the "Uranium Club" in a seminary held during the first years of the war. The conclusion was that if they worked on nuclear bombs at all, the work did not fully occupy them.

   b. *History will record that the Americans and the English made a bomb, and that at the same time the Germans, under the Hitler regime, produced a workable engine. In other words, the peaceful development of the uranium engine was made in Germany under the Hitler regime, whereas the Americans and the English developed this ghastly weapon of war.*
   This statement is sometimes quoted in the literature as proof that the German team congratulated itself for its "moral superiority" as compared to the American bomb builders. However, for CFvW it was just a statement of a paradoxical fact. When he made it he only knew that the Americans had made a Uranium bomb, and he did not know yet, before the Nagasaki bomb, that the Americans had also constructed reactors, ever since Fermi's first critical chain reaction already in December 1942, and had produced plutonium by operating large reactors. CFvW, as well as Heisenberg, never claimed "moral superiority" from the fact that they had not built atomic bombs. On the contrary, they expressed understanding for their American colleagues who were "on the good side" in the battle against Hitler and could have a good conscience in their – though hypothetical – race to let the democracies have the bomb before Hitler had it and used it for world domination.

**Final Remarks**

It may be hoped that the facts and arguments presented in this paper will help to reveal why the simplified saying "Heisenberg tried to build an atom bomb for Hitler and failed", mentioned in the Introductory Remarks above, and in various modifications still findable in contemporary texts, is based on historical misunderstandings, is misleading and, if a brief summary is needed, should be replaced by something like this: "Heisenberg, like several other German physicists, was drafted by German Army Ordnance when war began in Europe in September 1939 to investigate whether the energy from splitting Uranium nuclei by neutrons could be used for military purposes. Heisenberg found that this is possible in principle but would require such enormous industrial expenditures that it would take many years and would be impracticable while the war lasted. The project was therefore dropped by the Nazi Government in 1942. Heisenberg had even refrained from calculating a precise value for the critical mass of U 235. He was relieved that he was thus spared a moral decision between obeying an order to build the bomb or risking his life by refusing to be involved in the project or sabotaging it. He was happy to be confined to a project of building a small test reactor under civilian administration that the Government had approved. In 1941 Heisenberg had tried to get the opinion of Niels Bohr in Copenhagen on what the international community of nuclear physicists could possibly do or prevent regarding the long-range technical feasibility of making nuclear weapons. Bohr had misunderstood Heisenberg's cautious approach."

*This article appeared, in June 2016, on the website of the Heisenberg Society, [www.heisenberg-gesellschaft.de](www.heisenberg-gesellschaft.de). In a slightly modified form and under the title "Werner Heisenberg – Was He a Would-Be Pioneer of the Atom Bomb? The Longevity of Myths About Heisenberg's Activities During World War II" it was published in the July 2016 issue of the Physics and Society Newsletter of the American Physical Society.*

**Literature**

- Helmuth Albrecht – Armin Hermann, Die Kaiser-Wilhelm-Gesellschaft im Dritten Reich (1933 – 1945). In: Rudolf Vierhaus und Bernhard vom Brocke (Herausgeber) Forschung im Spannungsfeld von Politik und Gesellschaft. Geschichte und Struktur von Kaiser-Wilhelm-/Max-Planck-Gesellschaft, Deutsche Verlagsanstalt Stuttgart 1990, Seiten 397 – 399
- Jeremy Bernstein (Editor) Hitler's Uranium Club, The Secret Recordings at Farm Hall, with an introduction by David C. Cassidy, American Institute of Physics 1996.
- Bohr-Heisenberg Symposium Marks Broadway Opening of *Copenhagen.* In: Physics Today, May 2000, pages 51/52
- Carson, Cathryn. "Reflections on Copenhagen . In: Michael Frayn's Copenhagen in debate: Historical essays and documents on the 1941 meeting between Niels Bohr and Werner Heisenberg, ed. Matthias Dörries, Berkeley: Office for History of Science and Technology, 2005. Published in German as "Reflexionen zu 'Kopenhagen.'" In Michael Frayn, Kopenhagen: Mit zehn wissenschaftsgeschichtlichen Kommentaren, ed. Matthias Dörries, 3rd, rev.ed., 172-188. Göttingen: Wallstein, 2003. In initial form in 1st ed., 149-162. Göttingen: Wallstein, 2001

**Werner Heisenberg 2016**